# Effect of the Source-to-Substrate Distance on Structural, Optoelectronic, and Thermoelectric Properties of Zinc Sulfide Thin Films

Asad Ur Rehman Khan [1], Muhammad Ramzan [1,*], Muhammad Faisal Iqbal [2], Muhammad Hafeez [3], Mohammed M. Fadhali [4,5], Hamoud H. Somaily [6,7], Muhammad Javid [8], Muhammad Waqas Mukhtar [1] and Muhammad Farooq Saleem [9,*]

1. Institute of Physics, Baghdad-Ul-Jadeed Campus, The Islamia University of Bahawalpur, Bahawalpur 63100, Pakistan
2. Department of Physics, Riphah International University Faisalabad, Faisalabad 38000, Pakistan
3. Department of Physics, Lahore Garrison University, Lahore 54810, Pakistan
4. Department of Physics, Jazan University, Jazan 82812, Saudi Arabia
5. Department of Physics, Faculty of Science, Ibb University, Ibb 70270, Yemen
6. Research Center for Advanced Materials Science (RCAMS), King Khalid University, P.O. Box 9004, Abha 61413, Saudi Arabia
7. Department of Physics, Faculty of Science, King Khalid University, P.O. Box 9004, Abha 61421, Saudi Arabia
8. Institute of Advanced Magnetic Materials, College of Materials and Environmental Engineering, Hangzhou Dianzi University, Hangzhou 310012, China
9. GBA Branch of Aerospace Information Research Institute, Chinese Academy of Sciences, Guangzhou 51070, China
* Correspondence: mr.khawar@iub.edu.pk (M.R.); farooq@aircas.ac.cn (M.F.S.)

**Abstract:** Zinc sulfide (ZnS) thin films with variable structural, optical, electrical, and thermoelectric properties were obtained by changing the source-to-substrate (SSD) distance in the physical-vapor-thermal-coating (PVTC) system. The films crystallized into a zinc-blende cubic structure with (111) preferred orientation. The films had a wide 3.54 eV optical band gap. High-quality homogenous thin films were obtained at 60 mm SSD. The sheet resistance and resistivity of the films decreased from $10^{11}$ to $10^{10}$ Ω/Sq. and from $10^{6}$ to $10^{5}$ Ω-cm, when SSD was increased from 20 mm to 60 mm, respectively. The phase and band gap were also verified by first principles that were in agreement with the experimental results. Thermoelectric characteristics were studied by using the semi-classical Boltzmann transport theory. The high quality, wide band gap, and reduced electrical resistance make ZnS a suitable candidate for the window layer in solar cells.

**Keywords:** ZnS thin films; physical-vapor-thermal-coating technique; source-to-substrate distance; solar cells

## 1. Introduction

Serious challenges such as the energy crisis and pollution due to the huge rise in chemical exhaust as well as the burning of fossil fuels, including coal, have motivated scientists and researchers to search for better energy alternatives. Solar energy is a better alternative source of energy that is renewable as well as clean and inexhaustible [1]. To expedite the growth of the optoelectronics industry, especially for solar cells, it is necessary to boost the efficiency of thin-film devices as well as to reduce the production cost. Thin-film technology is one of the bases for high-efficiency, environmentally friendly, and low-cost devices. Several factors such as doping, surface morphology, plasma treatments, and deposition techniques contribute to improving the performance of nanostructured thin films [2]. For optoelectronics applications, thin films of III–V gallium arsenide (GaAs),

IV–VI lead sulfide (PbS), and II–VI compounds consisting of cadmium sulfide (CdS), zinc sulfide (ZnS), cadmium telluride (CdTe), zinc telluride (ZnTe), cadmium selenide (CdSe), zinc selenide (ZnSe), etc., are widely used [3–6]. Among these, the II–VI semiconductors have numerous applications in the optoelectronics industry such as light-emitting devices (LEDs), photodetectors, small integrated circuits, catalysis, energy generation, antireflection coatings, laser screens, thin film transistors (TFT), and ultrasonic transducers, due to tunable physical and unique optical properties [7,8]. CdS is generally used as a window and buffer layer in CdTe and copper-indium-gallium-selenide (CIGS) solar cells, respectively. CdS is toxic to the environment and it has a narrow optical bandgap (~2.42 eV). These drawbacks of CdS thin films are opening the window for researchers to find suitable alternatives for optoelectronics applications. Remarkable improvements have been made to develop new electronic devices with good optical transparency, thermal stability, and transport properties. Currently, ZnS thin films are fascinating researchers due to their optoelectronic applications such as: optical filters, LEDs, gas sensors, and window materials for thin film solar cells [9,10]. ZnS thin films exhibit high optical transmittance, high refractive index (2.35), wide bandgap (3.7 eV), non-toxicity [11], and are in large-scale production [12]. The optical bandgap of ZnS thin films lies in the range of 3.74 to 3.91 eV for hexagonal structure and 3.54 to 3.72 eV for cubic structure [13]. In comparison with CdS thin films, ZnS films are environmentally friendly with better optical properties and can be a suitable replacement in window and buffer coatings in thin film solar cells. ZnS can improve the efficiency of solar cells as well as improve the sunlight transmittance in the buffer layer. Moreover, in comparison to CdS, ZnS shows better lattice matching to the absorber layer in CIGS solar cells [14]. The higher optical absorption for wavelengths < 520 nm makes it suitable for various photovoltaic applications [15]. ZnS can be deposited on various types of substrates and devices such as glass, silicon, GaAs, and ITO for various applications [16]. There are number of chemical and physical routes for the deposition of ZnS thin films such as the ultrasonic spray technique [17,18], the sol–gel process [19], thermal evaporation [20], electrochemical deposition [16], electron beam evaporation [21,22], the spray pyrolysis technique [23,24], the SILAR method [25], rf-magnetron sputtering [26], atomic layer deposition [27], and chemical bath deposition [14,28].

Deposition of ZnS thin films for solar cell applications by using thermal evaporation is always a challenge because the growth rate of thermal evaporation is approximately 1 µm/h, which can be controlled by the deposition conditions, but for solar cells applications the window layer must be close to 100 nm to achieve optical transmittance [29]. We successfully deposited ZnS thin films by the simple and cost-effective physical vapor thermal coating (PVTC) technique. The focus of the current study was divided into three parts. Firstly, the effect of SSD on the structural, optical, and electrical properties was investigated in the evaporated ZnS thin films. Secondly, the optoelectronic and thermoelectric properties of the films were theoretically studied. In the third part, we correlated the experimental and theoretical results for solar cell applications.

## 2. Experimental and Computed Methods

ZnS (99.99%) in powder form was procured by Sigma Aldrich and used for the deposition on glass substrates without any further purification. The glass substrates were first treated with detergents and well-washed with running water. Ethanol was used to clean the glass substrates after they were dried in a microwave oven at 90 °C. A molybdenum boat was used to hold the ZnS powder and substrates were placed on the substrate holder. Diffusion and rotatory pumps attached to the deposition system were used to create the base vacuum level of ≤$10^{-5}$ Torr and the deposition pressure was ≤$10^{-4}$ Torr inside the working chamber. On achieving the desired vacuum, a 300 A evaporation current was applied to the source holding the boat. ZnS was evaporated on a glass substrate. The total deposition process took place in 10 min. Conventionally, the SSD is the distance between the source-holding boat and the substrate holder, which is usually a few millimeters. We varied the SSD from 20 to 80 mm and developed four samples, namely, "ZnS-A" at 20 mm, "ZnS-B" at 40 mm, "ZnS-C" at 60 mm, and "ZnS-D" at 80 mm.

The structure of the films was examined by using a D8 advance Bruker, Cu-Ka (1.54 Å) X-ray diffractometer (XRD). The surface morphology was analyzed by a Hitachi SU-70 scanning electron microscope (SEM). Optical analysis was carried out by using a Lambda 25, Perkin Elmer UV-Vis dual beam spectrophotometer. Structural, morphological and optical analysis experimentally performed in Central Hi-Tech Lab (CHL), Government College University Faisalabad (GCUF), Pakistan. The electrical properties of the films were measured by the locally built four-probe-point (4PP) method at room temperature. Electrical analysis experimentally performed in Centre for Advanced Electronics and Photovoltaic Engineering (CAEPE), International Islamic University Islamabad (IIUI), Pakistan.

This investigation [30] uses the Wien2k algorithm, using the DFT-based full-potential linearized augmented-plane-wave (FP-LAPW) approach [31]. The most precise semi-local exchange potential (mBJ) modification approach was used to calculate the electrical and optical properties [32]. BoltzTraP code was used to calculate the thermoelectric properties.

The coefficient of electrical conductivity $\sigma_{\alpha\beta}(\varepsilon)$ is given below.

$$\sigma_{\alpha\beta}(\varepsilon) = \frac{1}{N}\sum_{i,k}\sigma_{\alpha\beta}(i,k)\frac{\delta(\varepsilon - \varepsilon_{i,k})}{\delta(\varepsilon)} \quad (1)$$

$$\sigma_{\alpha\beta}(i,k) = e^2\tau_{i,k}v_\alpha(i,\vec{k})v_\beta(i,\vec{k}) \quad (2)$$

where $\tau$ is the relaxation time, and $v_\beta$ (*i*, *k*) is the group velocity of the parameters that were executed in the BoltzTraP code.

Various comparisons [33–36] show that for the measurement of band gaps the mBJ is the best semi-local approximation at a lower computational price, receiving an average accuracy even better than that of one of hybrid functionals [36]. For larger and smaller band gaps in very heterogeneous systems, the mBJ approach is used, while PBE systematically underestimates local band gaps.

## 3. Results and Discussions

### 3.1. Structural Analysis

ZnS thin films can be deposited in hexagonal and zinc blend (cubic) structures, depending upon the deposition technique and parameters [17]. The hexagonal structure is stable at atmospheric pressure and higher temperatures, while at room temperature cubic structure is stable [37]. Generally, ZnS thin films developed by the RF sputtered technique have a hexagonal phase [38], those developed by the spray pyrolysis method have a mixed phase [39], and those developed by thermal evaporation, a cubic phase [40,41]. The cubic to hexagonal transition of phase takes place at around about 1313 K [42]. Figure 1, depicts the XRD pattern of ZnS thin films deposited by the PVTC method by varying the source-to-substrate distance from 20 mm to 80 mm. In all four samples (ZnS A–D), all diffraction peaks (111), (220), and (311) observed at ~28.56°, 47.52°, and 56.29°, correspond to the zinc blende cubic phase of ZnS with preferred orientation along (111) plane, (JCPDS No. 25-2566). Similar crystal structure and cubic growth orientation along the (111) plane are reported by Xue [43], Jin et al. [44] and Zhang et al. [45]. The (220) and (311) planes are also observed in XRD patterns confirming the zinc blend structure of ZnS. The presence of more than one crystal plane in the XRD pattern confirms the polycrystalline nature of the film. No impurity-related extra peak is observed in the XRD patterns of samples A–D, showing the high purity of the samples. ZnS nanostructures can be developed easily in the wurtzite (hexagonal) or wurtzite–cubic mixed phase as compared to the single cubic phase [46,47]. In this work, it is evident that there is no crystal plane related to the hexagonal structure so deposited films are cubic in nature. Figure 1 illustrates that the intensity of the (111) peak increased with increasing SSD from 20 to 60 mm and after that, it decreased for 80 mm SSD. Given the same deposition time (10 min), the increase in (111) peak intensity with the increase in SSD up to 60 mm and the decline when the SSD was further increased to 80 mm, can be described in terms of the energies of the flux of depositing species approaching the substrate. For the ZnS thin films deposited at 60 mm SSD, the energies of deposited species allow them to organize in the (111) direction, resulting in the maximum diffraction and hence the maximum peak intensity (111). The decline in the diffraction intensity of the (111) crystal plane with further increase in SSD can be attributed to the lack of energy of deposited species to reach the substrate (placed at 80 mm) and organize in the (111) crystal plane [48]. The high peak intensity for 60 mm SSD may be due to the slow nucleation as compared to 20, 40, and 80 SSD. Slow nucleation improves the growth rate; as the growth rate increases the size of crystallites increases. Moreover, peak intensities of the (220) and (311) planes were observed to be decreased with increasing SSD and to almost disappear at 80 mm SSD. The average crystallite size (D) for (111) planes was estimated by using the Scherrer equation [49], as follows:

$$D = \frac{K\lambda}{\beta cos\theta} \qquad (3)$$

where $K$ is a dimensionless constant with a value that lies between 0.9 to 1.0, $\beta$ in radian provides the full width half maximum (FWHM), $\lambda$ represents the wavelength (Cu-K$\alpha$ radiation, 1.5406 Å), and $\theta$ is the angle of diffraction. The average crystallite size of the films increased from 26 to 29 nm for 20 to 60 mm SSD and decreased to 27 nm for 80 mm SSD.

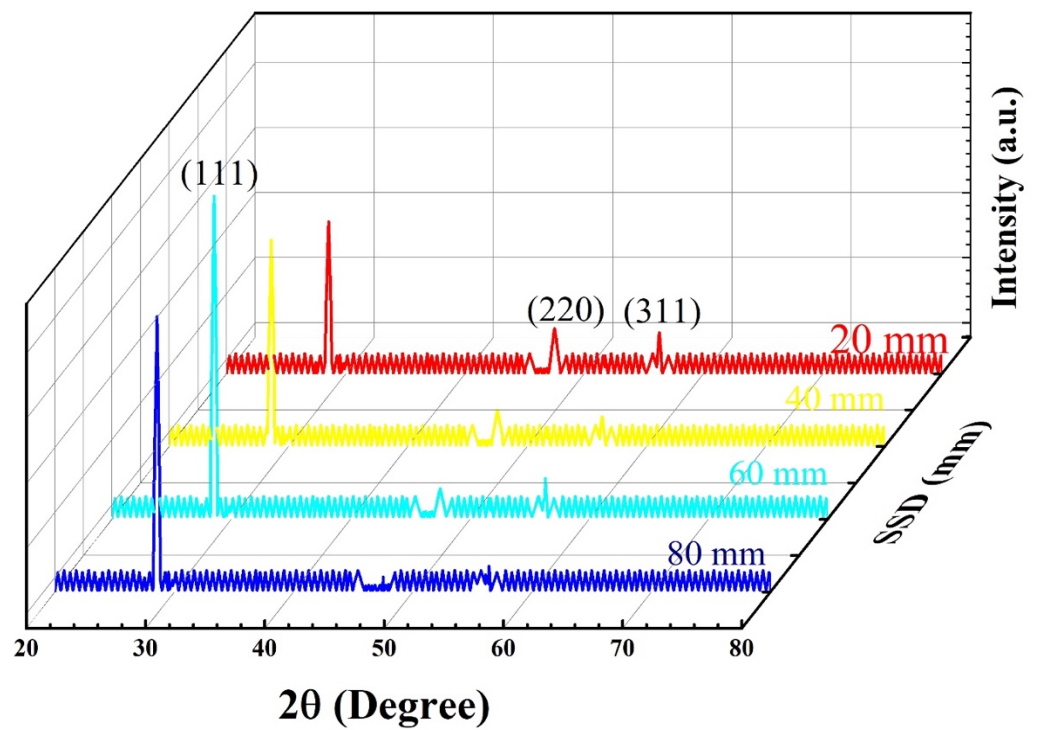

**Figure 1.** XRD spectrum of ZnS thin films deposited at various SSD (mm).

Figure 2a shows the variation of average crystallite size '$D$' of ZnS films and '$\beta$', full width at half maximum (FWHM) of (111) planes, as a function of SSD. Dislocation density "$\delta$" and microstrain "$\varepsilon$" were calculated for the (111) peak by using the following Equations (4) and (5) [15].

$$\delta = \frac{1}{D^2} \qquad (4)$$

$$\varepsilon = \frac{\beta cos\theta}{4} \qquad (5)$$

The lattice constant "*a*" of the zinc blend structure of the films was measured by the following Equation (6) [49].

$$\frac{1}{d_{hkl}^2} = \frac{h^2 + k^2 + l^2}{a^2} \tag{6}$$

where "$d_{hkl}$" is the interplanar spacing, measured ~0.3121 nm by using Bragg's Law [50].

The measured average value of "*a*" was 0.5407 nm which is in strong agreement with the literature for bulk ZnS nanostructures [51].

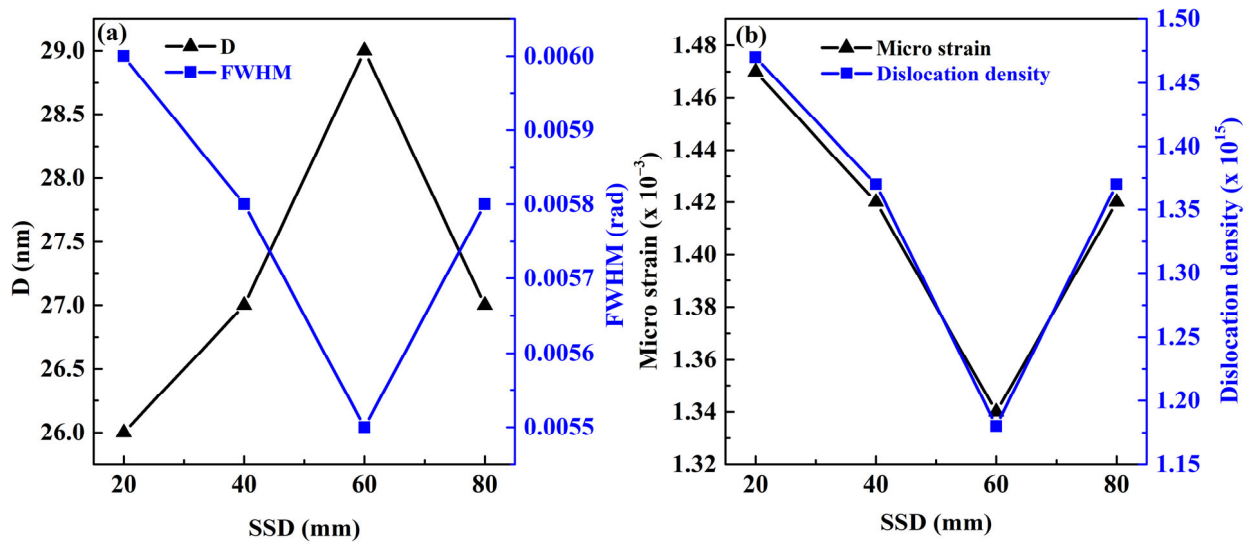

**Figure 2.** (**a**) The FWHM (rad) of the (111) plane and crystallite size (nm) and (**b**) the microstrain (×10$^{-3}$) and dislocation density (×10$^{15}$ m$^{-2}$) of deposited ZnS thin films against SSD.

The variation of microstrain "ε" and dislocation density "δ" of the films with increasing SSD from 20 to 80 mm are plotted in Figure 2b. It was observed that micro strain "ε" and dislocation density "δ" decreased from 1.47 × 10$^{-3}$ to 1.34 × 10$^{-3}$ and 1.47 × 10$^{15}$ to 1.18 × 10$^{15}$ m$^{-2}$, respectively, with increasing SSD from 20 to 60 mm. With further increase in SSD from 60 to 80 mm, the "ε" and "δ" increased up to 1.42 × 10$^{-3}$ and 1.37 × 10$^{15}$ m$^{-2}$, respectively. Full width half maximum "β" decreased from 0.0060 to 0.0055 (rad) with increasing SSD from 20 to 60 mm, which caused the improvement in crystallite size "D" [52], reduction in microstrain "ε", and dislocation density "δ". It can be observed that the value of "δ" was 1.18 × 10$^{15}$ m$^{-2}$ of the film at 60 mm SSD because grain size and dislocation density have an inverse relationship as confirmed by Equation (4). The increase in microstrain "ε" may have been due to the increment in lattice parameter and the absence of vacancies and defects [51]. The values of structural parameters as a function of SSD are given in Table 1.

**Table 1.** Structural parameters by XRD data and optical bandgaps by Tauc's plot method were obtained.

| Samples ID | ZnS-A | ZnS-B | ZnS-C | ZnS-D | References |
|---|---|---|---|---|---|
| SSD (mm) | 20 | 40 | 60 | 80 | |
| FWHM (rad) | 0.0060 | 0.0058 | 0.0055 | 0.0058 | 0.005, 0.0053 [20], 0.0096 [22] |
| Crystallite size "D" (nm) | 26 | 27 | 29 | 27 | 15 [22], 29.6, 28.2 [20], 38 [17] |
| Micro strain "ε" (×10$^{-3}$) | 1.47 | 1.42 | 1.34 | 1.42 | 1.22, 1.28 [20], 0.913 [17], 0.23 [22], 8.4 [53] |
| Dislocation density "δ" (×10$^{15}$ m$^{-2}$) | 1.47 | 1.37 | 1.34 | 1.37 | 0.69 [17], 4.1 [22] |
| Band gap (eV) | 3.31 | 3.50 | 3.54 | 3.47 | 3.48 [17], 3.46 [22], 3.75 [54], 3.5 [53] |

*3.2. Surface Morphology*

Surface morphology of the films was revealed by SEM micrographs. In Figure 3a–d, it can be seen that the nanoparticles are unevenly distributed and agglomeration has occurred over the surface of substrates, while nanoparticles have good connectivity among them. Moreover, in ZnS thin films A, B, and D, formation of flower-like nanoclusters occurred. The thin film "A" and "B" contain spikes, while the structure of "D" thin film is a combination of flake-like nanoclusters. The thin film "C" deposited at 60 mm SSD has relatively smooth coverage of homogeneity over the substrate and is compact with almost uniform visible grain sizes. SEM micrographs also show that all the particles have sizes in nanometers, which are in accordance with XRD results. The thicknesses of the films are ~219, 197, 186, and 191 nm for ZnS-A, ZnS-B, ZnS-C, and ZnS-D, respectively. The thickness of the films reduced from 219 nm to 186 nm which could have been due to the mass-limited reaction kinetics that caused the increment in the rate of evaporation and the further increase in the thickness to 191 nm could have been due to improvements in reaction kinetics on the surface of the substrate that produces the nucleation sites [55]. In Figure 3a,b,d), thick dark fringes among grains may not be grain boundaries, these could be voids because the films are ultra-thin layers. These undesirable voids are not corporate (as an n-type buffer layer) in the efficiency of thin film solar cells because the shunting effect can take place [56]. Localized and non-homogeneous heterojunctions with p-type absorber layers (CIGS, CZTSSe) occur due to the non-uniform ZnS buffer layer, which causes potential spatial distribution that does not support the drift of photo-generated carriers towards back and front terminals [56]. Figure 3c shows the nano-sized crystallized grains with pores for the ZnS film deposited at 60 mm SSD where the particles around the pore and the pore walls are of fused grains. In Figure 3, insets show that the ZnS thin films have good adherence to substrates.

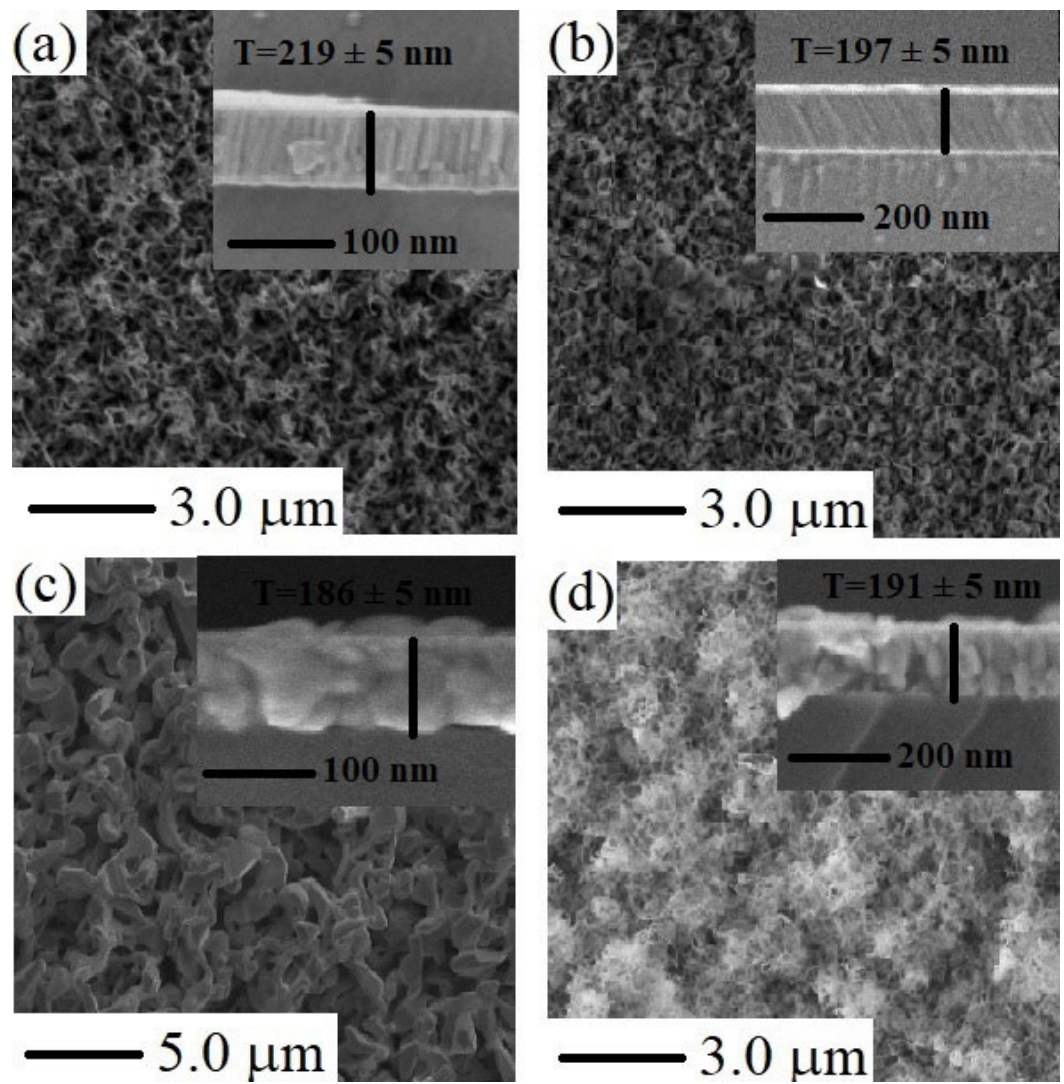

**Figure 3.** SEM images of ZnS thin films deposited at (**a**) 20 mm, (**b**) 40 mm, (**c**) 60 mm, and (**d**) 80 mm SSD. Insets show the cross-sectional zoom in micrographs of ZnS thin films for thickness measurement.

*3.3. Optical Analysis*

Figure 4a,b show the typical optical transmittance and absorbance spectrum of the films at various SSD. The films show higher optical transmission in the longer wavelength range and much lower optical transmittance in lower wavelength regions. The films deposited at 40 to 80 mm SSD, show an average optical transmittance (T) > 65% in the visible region. The average optical transmittance increases and optical absorbance reduces with increasing SSD from 20 to 60 mm, and then reduces to 80 mm. The 186 nm thick film deposited at 60 mm SSD shows the maximum average optical transmittance of ~76% and minimum average optical absorbance of ~13% in the visible region. The enhancement in optical transmittance with increasing SSD from 20 to 60 mm is due to the reduction in thickness (the lower the thickness the higher will be the optical transmittance); ZnS-C is the thinnest film as compared to other samples and it has higher optical transmittance as compared with other films. This could be due to the reduction in scattering and defects, which increases the optical transmittance [57]. Improvement in optical transmittance could also be due to the changing of the non-homogenous structure to an homogenous one as illustrated in the SEM result, because the scattering of light is due to an irregular arrangement of atoms. The average optical absorbance of all the films is in the range of ~13–42% and it reduces with increasing SSD.

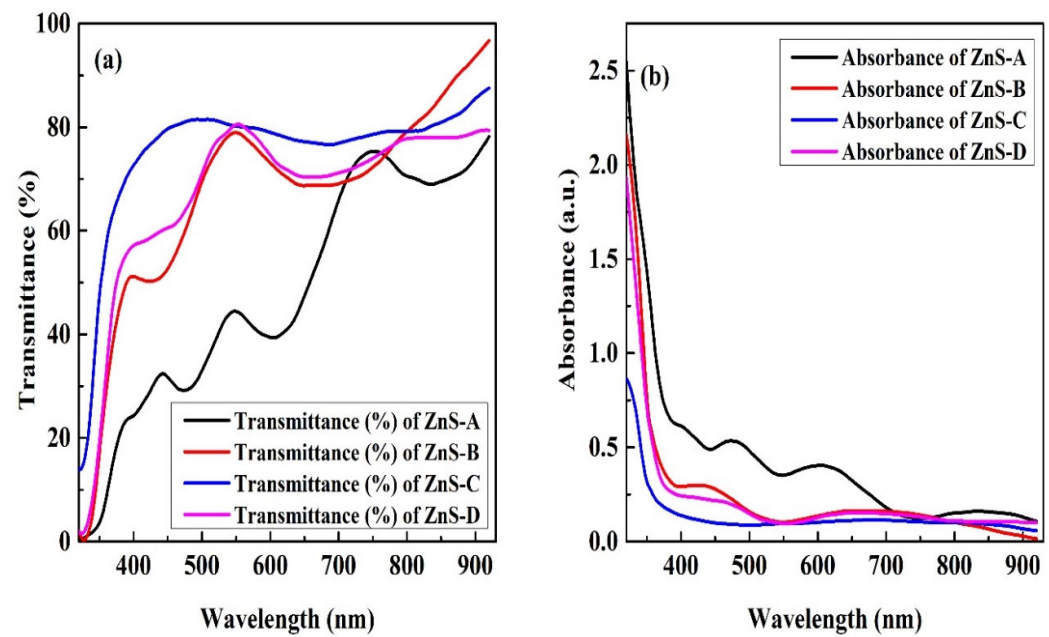

**Figure 4.** (**a**) Optical transmittance and (**b**) absorbance of deposited ZnS thin films.

The optical band gaps of all deposited samples were measured by Tauc's relationship. The extrapolating of the linear portion of $h\upsilon$ versus $(\alpha h\upsilon)^2$ curve provided the optical band gap of samples [58].

$$\alpha h\upsilon = A(h\upsilon - E_g)^n \tag{7}$$

where $A$ is the band tailing parameter, the slope of the Tauc line; $n$ values depend on the transition probability, 2 for directly allowed band gap; $h\upsilon$ (eV) is photon energy; $\alpha$ is the absorption coefficient; and $E_g$ is the optical band gap of the material. The $E_g$ increases from 3.31 to 3.54 eV with increasing SSD from 20 mm to 60 mm and further reduces to 3.47 eV for 80 mm SSD, as shown in Figure 5a–d. The surface morphology clearly changed with SSD, as shown in Figure 3. The shifting of the band gap strongly depends upon the morphology and structure of the films [59,60]. The band shifting and other all properties of the films are influenced by the structure and surface morphology of the films. The widening of "$E_g$" with increasing SSD up to 60 mm might have been due to crystallinity of the films supported by SEM and XRD analysis. The growth of crystallites was influenced by SSD which caused the widening of "$E_g$" that can be due to the Burnstein–Moss band shift [23]. The direct band gap "$E_g$" decreases from 3.54 to 3.47 eV when SSD increases from 60 to 80 mm which might have been due to the chances of more realignment and oxidation, and better interaction between substrate and deposited layer [21,61]. The relation among molar reflectivity, molar volume, and refractive indices "$n$" is estimated by the following Equations (8) and (9) [62].

$$\frac{R_m}{V_m} = \frac{n^2 - 1}{n^2 + 1} \tag{8}$$

$$n = \frac{1 + \sqrt{R}}{1 - \sqrt{R}} \tag{9}$$

where $R$, $Rm$, and $Vm$ are the optical reflectance, molar reflectivity, and molar volume, respectively.

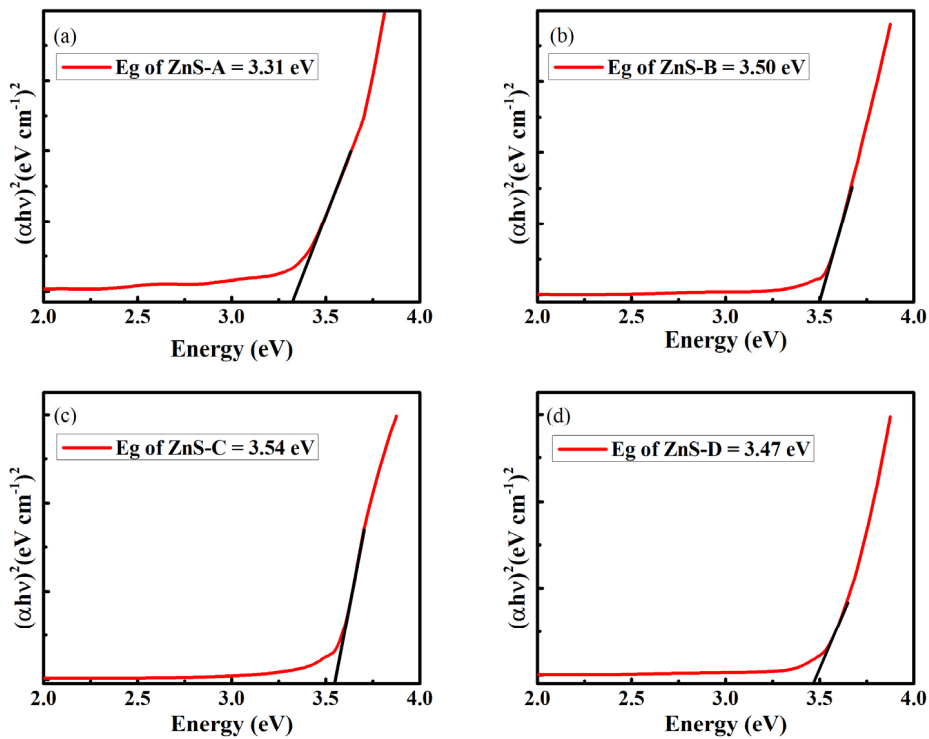

**Figure 5.** The optical band gap of (**a**) ZnS-A, (**b**) ZnS-B, (**c**) ZnS-C, and (**d**) ZnS-D thin films.

Figure 6a shows the refractive indices (*n*) for all the films deposited at various SSD with the wavelength in the region 380–800 nm measured by using the optical reflectance "*R*". The graphical results indicate the high values of refractive index in the Vis region. Moreover, in the visible range, the refractive indices have values between ~1.55 and 2.1 for all samples and these values are favorable to the transmission of light in the desired region [63]. The behavior of the ratio *Rm/Vm* of all the films is illustrated in Figure 6b. The ratio indicates the overall decreasing variation in the 380–800 nm wavelength region for all samples except the ZnS thin film deposited at 20 SSD. *Rm/Vm* and "*n*" decreased with increasing the SSD up to 60 mm.

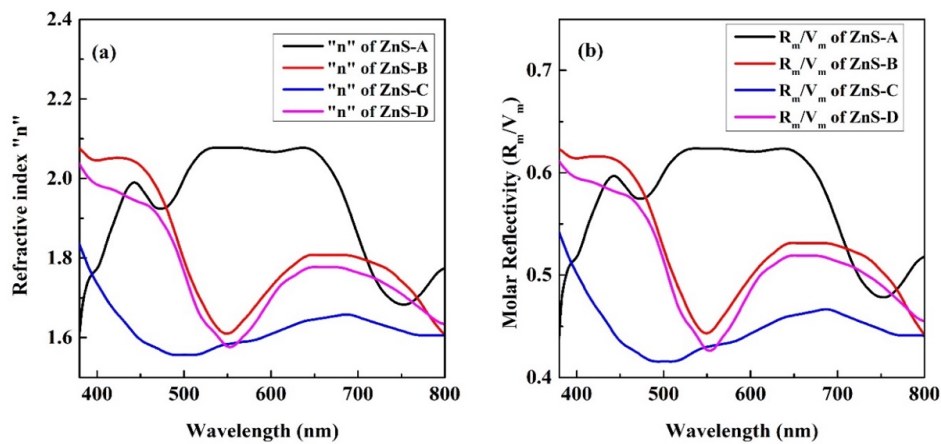

**Figure 6.** (**a**) Refractive indices and (**b**) molar reflectivity per unit volume are shown for ZnS deposited films against wavelength.

The real and imaginary parts of the dielectric constant $\varepsilon_1(\omega)$ and $\varepsilon_2(\omega)$, respectively, are used to define the optical characteristics and are drawn in Figure 7a–f. The linkage of $\varepsilon_1(\omega)$ and $\varepsilon_2(\omega)$ can be described by the Kramer–Kronig equation [64].

$$\varepsilon(\omega) = \varepsilon1(\omega) + i\varepsilon2(\omega) \tag{10}$$

Figure 7a illustrates the real dielectric portion that results from the dispersion of incoming light from any material at plasmonic frequency. Two prominent peaks are observed at 5.3 eV and 6.5 eV, respectively, where the ions are aligned to fully utilize the polarization and the value for $\varepsilon_1(\omega)$ rises from the static limit, say $\varepsilon_1(0)$. The dielectric constant's imaginary component, $\varepsilon_2(\omega)$, describes the quantity of light absorbed by the material. In the case of ZnS films, from the static limit $\varepsilon_1(0)$, the value of $\varepsilon_2(\omega)$ rises until it reaches the highest peak values at 6.6 eV. The dielectric loss factor (tan(δ)) is the ratio of the imaginary part of the dielectric to the real part of the dielectric, which is represented in Figure 7g. The dielectric loss factor gives the measurement of absorbed energy in the medium when an electromagnetic wave transmits through the medium. In the ideal case, the dielectric loss factor and losses are zero. The dielectric loss of ZnS has a peak value at 7.4 eV due to the contribution of electronic, ionic, dipolar, and space charge polarization [65]. The real part of the dielectric constant $\varepsilon_1(\omega)$, the imaginary part of the dielectric constant $\varepsilon_2(\omega)$, the extinction coefficient '$k(\omega)$', and the refractive index '$n(\omega)$' are linked as, $n^2 - k^2 = \varepsilon_1(\omega)$ and $2nk = \varepsilon_2(\omega)$. The $\varepsilon_1(\omega)$ and $n(\omega)$ begin to increase from their critical limits and approaches, both at almost similar peak energies.

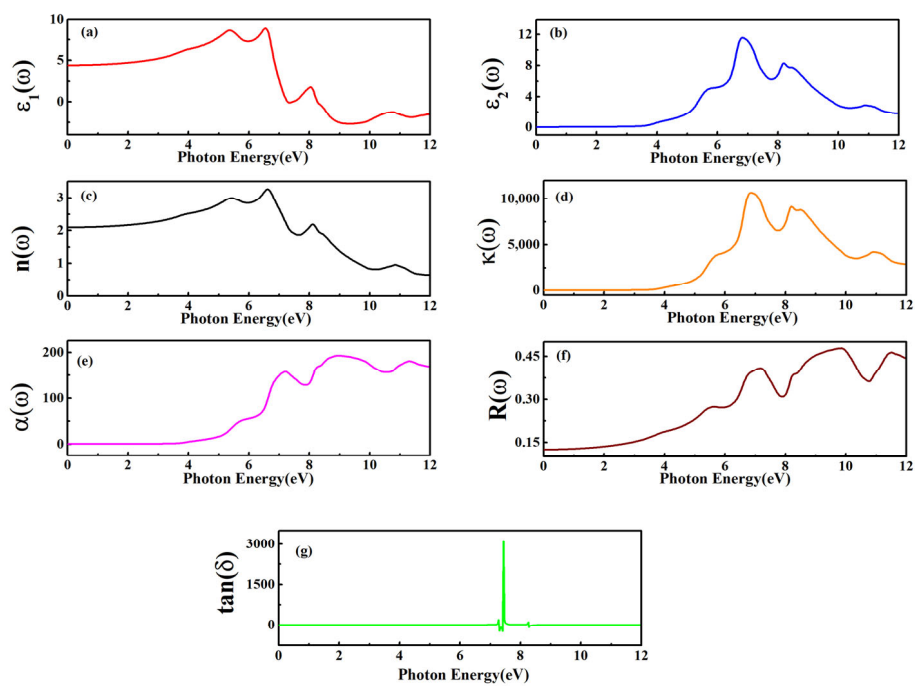

**Figure 7.** Computed values of (**a**) ε1(ω), (**b**) ε2(ω), (**c**) n(ω), (**d**) k(ω), (**e**) α(ω), (**f**) R(ω), and (**g**) tan (δ) for ZnS.

The frequency-dependent variation of *k(ω)* is shown in Figure 7d, describing the response of ZnS films exposed to light of varying densities. From Beer's law, the light attenuation is defined by the relationship between absorption "*A*", extinction coefficient (or molar absorptivity) "*k*", thickness of the sample "b", and the ratio of the number of atoms in unit volume for the bulk materials "c" as A = kbc. Moreover, the static value calculations of $\varepsilon_1(0)$ and n(0) exactly follow the mathematical expression n(0)² = $\varepsilon_1(0)$. Another important parameter that shows the quantity of light absorption is the absorption coefficient α(ω), which categorizes the type of photon as well as its energy. A maximum absorption peak of 157 cm at 7.17 eV was found, as shown in Figure 7e. The surface morphology elaborates on how light reflects. The energy area due to maximum reflection develops the real part of the dielectric constant, where the maximum reflection occurs, as illustrated in Figure 7c. Lower values of reflectivity R(ω) and larger values of absorption coefficient α(ω) for ZnS are observed in Figure 7e,f. The compounds with metallic character have stronger visible-light absorption capability and improved photovoltaic properties. Meanwhile, the mean peaks of absorption are found in the UV region at an energy range of ~5–10.5 eV for all studied compounds [66]. When the absorbed energy is higher, the electron-hole pair for conduction is larger. ZnS is a suitable candidate for solar applications since its optical properties reveal that it has high absorption and the least amount of reflection.

*3.4. Electrical and Electronic Analysis*

The four-point-probe (4PP) technique was used to measure the sheet resistance and resistivity by using the following Equations (11) and (12) [63].

$$\rho = \frac{\pi}{\ln(2)} * \frac{V}{I} \quad (11)$$

$$R = \rho T \quad (12)$$

where "*I*" and "*V*" represent the applied current between the outer probes and the voltage difference between inner probes, respectively. Here "ρ" represents the sheet resistance in Ω/Sq, "*R*" is the electrical resistivity in Ω-cm, and "*T*" is the thickness of the films.

Figure 8 shows the voltage and current plot as a function of SSD and the measured sheet resistances of the films are given in Table 2. The sheet resistance and resistivity of the films decreased from $3.14 \times 10^{11}$ to $1.54 \times 10^{10}$ Ω/Sq. and from $6.89 \times 10^6$ to $2.87 \times 10^5$ Ω-cm, with an increase in SSD from 20 to 60 mm, respectively. Many factors, such as crystallite size, crystallite structure, carrier mobility, and carrier concentration, have a crucial role in defining the electrical properties of the films. The reduction in electrical resistivity up to 60 mm SSD is attributed to the improvement in the crystallinity of the films as illustrated by XRD and SEM analysis. The increase in grain size reduces the grain boundary scattering, which contributes to the reduction in the number of trapping states for the electrons and this allows easy transportation of charge carriers in the ZnS lattice. It is observed that the measured electrical resistivities of the films are lower than those reported by Derbali et al. [17], who studied the electrical resistivity of ZnS thin films deposited by the spray ultrasonic technique. The reduction in resistivity is due to improvement in crystallinity but vacancies, interstitials, etc., are the other factors that can cause the increment in resistivity [67]. An increase in grain size and improvement to lesser lattice defects observed by SEM from 20 to 60 mm SSD (as shown in Figure 3) are responsible for the reduction of resistivity.

The valance band maximum (VBM) and conduction band minimum (CBM) in Figure 9 specifies that the outcomes reside at the zone center (Γ point) for ZnS, suggesting its direct band gap.

**Table 2.** Sheet resistance values for deposited ZnS thin films at different source-to-substrate distances.

| Sample ID | Sheet Resistance (×10¹⁰ Ω/Sq.) | Electrical Resistivity (×10⁵ Ω-cm) |
|---|---|---|
| ZnS-A | 31.4 | 68.9 |
| ZnS-B | 20.6 | 40.5 |
| ZnS-C | 1.54 | 2.87 |
| ZnS-D | 1.71 | 3.25 |

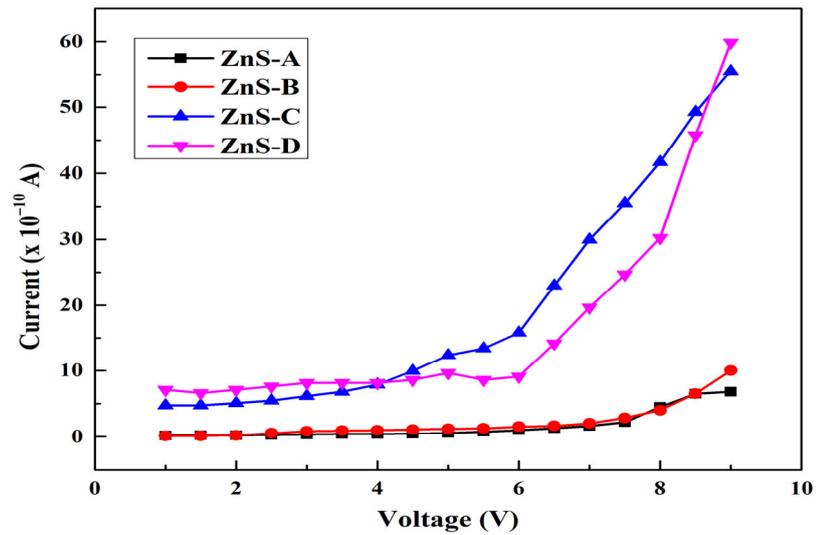

**Figure 8.** Voltage and current plot of ZnS-A, ZnS-B, ZnS-C, and ZnS-D thin films.

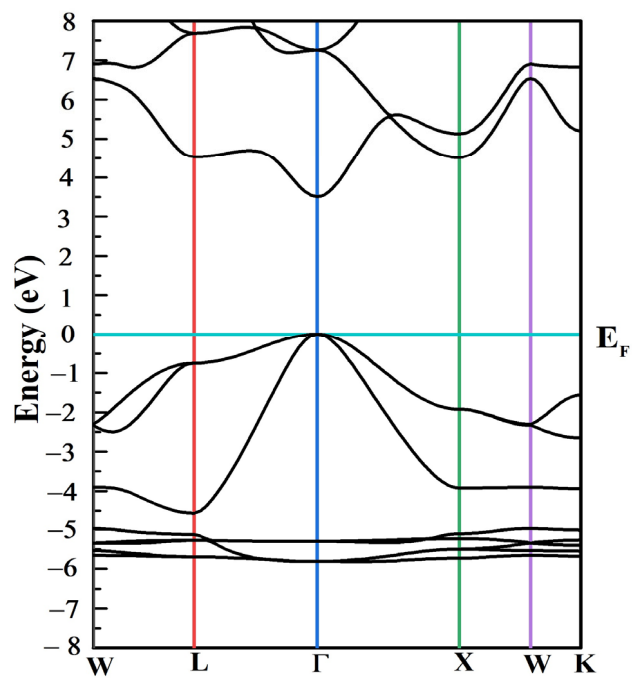

**Figure 9.** Computed electronic band structure of ZnS.

*3.5. Thermoelectric Properties*

Figure 10a–d elaborate the thermoelectric performance of ZnS as a function of temperature (T) ranging from 200 to 800 K using electrical conductivity ($\sigma/\tau$), thermal conductivity ($\kappa/\tau$), the Seebeck coefficient (S), and figure of merit (ZT). High $\sigma/\tau$ values and a higher S support improved thermoelectric efficiency.

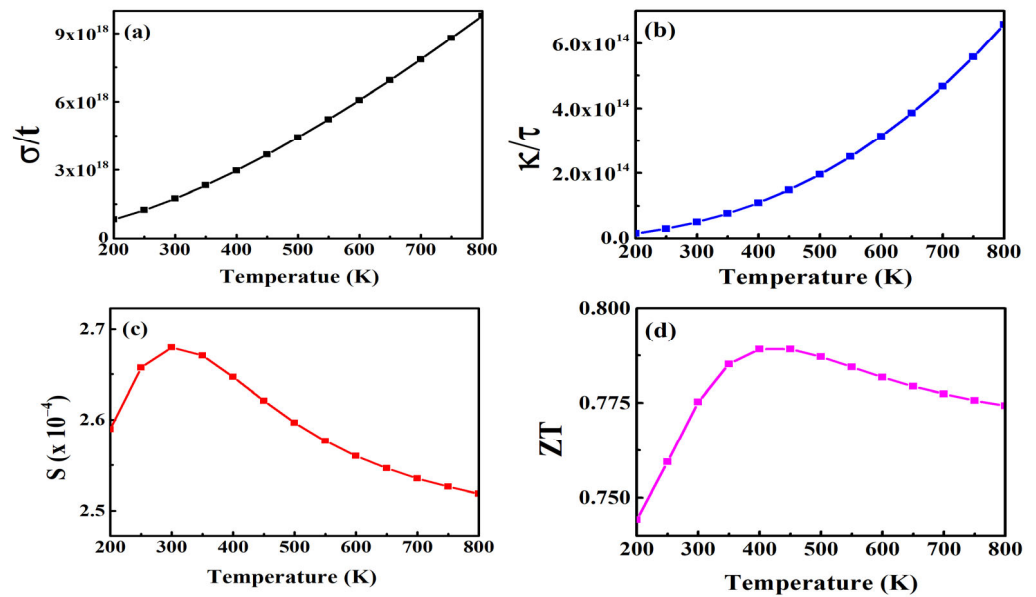

**Figure 10.** Computed values of (**a**) σ/τ, (**b**) κ/τ, (**c**) S, and (**d**) ZT plotted against temperature (300–800) K for ZnS.

According to Figure 10a, the value of σ/τ rises linearly with temperature in the region of 200 to 800 K. The electrical conductivity (σ/τ) in Figure 10a linearly increases with the temperature. The increase in temperature causes lattice vibrations that are the leading source of elastic waves for phonons' contribution, while free carriers are the source of the electrons' contribution. The plot of κ/τ versus T presented in Figure 10b shows linearly increasing behavior with respect to temperature. The κ/τ is a measurement of the thermal current generated by the electrons and phonons energy in the crystal structure. The ShengBTE code is typically used to calculate the phononic portion (Figure 10b). The relationship between the difference in voltage and the temperature difference is described by the S as S = V/T. Thermopower is another name for the Seebeck coefficient [68]. Depending on the type of charge, such as positive for holes and negative for electrons, it may have a positive or negative value [69]. The computed S increases at room temperature and begins to decrease with increasing temperature, as shown in Figure 10c. The dimensionless figure of merit is represented in Figure 10d and is given by relation ZT = S$^2$/T. ZnS has a ZT of 0.77 at room temperature, as seen in Table 3. It is typically used to describe a thermoelectric material's overall performance. The ratio of thermal to electrical conductivity is given by expression as LT = κ/σ [70] which states the Wiedemann–Franz law. The κ/σ ratio in these computed compounds is 10$^{-5}$, which makes them promising for thermoelectric applications. The thermoelectric properties are used at room temperature to fabricate suitable practical devices.

**Table 3.** Computed optical parameters, $a_o$ (Å) lattice constant, $E_g$, and thermoelectric coefficients of ZnS at room temperature.

| Properties | Computed Values | Reference Values |
|---|---|---|
| $\varepsilon_1(0)$ | 4.39 | ------ |
| $n(0)$ | 1.68 | ------ |
| R(0) | 0.01 | -------- |
| $a(\omega)$ | 7.17 eV, 8.97 eV | 5–10.5 eV [66] |
| $a_o$ (Å) | 5.40 | 5.44 [71], 5.45 [72] |
| $E_g$ (eV) | 3.47 | 3.71 [73], 2.09, 3.68 [74] |
| $\sigma/\tau$ (Ωms)$^{-1}$ | $1.68 \times 10^{18}$ | ------ |
| $\kappa/\tau$ (W/mKs) | $4.95 \times 10^{13}$ | ------- |
| (S) (V/K) | $2.72 \times 10^{-4}$ | ------- |
| ZT | 0.77 | ------- |

## 4. Conclusions

ZnS thin films made by changing the SSD were successfully deposited on glass substrates by using the PVTC technique. The films showed the zinc-blende or cubic structure with preferred orientation along the (111) plane. The 186 nm thick film developed at 60 mm SSD achieved over 76% optical transmittance in the visible region. The measured $E_g$ values of the films were in the range of 3.31 to 3.54 eV. The film deposited at 60 mm SSD had better crystallinity, void-free surface, higher optical transmittance, and low sheet resistance, and, due to this, it could be a better choice for solar cell applications such as a window or buffer layer.

The absorption coefficient and complex dielectric function favor the use of ZnS in different solar applications. In our studied compounds the figure of merit was 0.77, which means that these materials have potential applications as thermoelectric materials. The simulation results propose directions toward various applications of ZnS in the optoelectronic and thermoelectric industries.

**Author Contributions:** Conceptualization, A.U.R.K.; Data curation, M.H., M.M.F., H.H.S., M.J., M.F.S. and M.F.I.; Formal analysis, A.U.R.K., M.J., M.F.S. and M.F.I.; Funding acquisition, M.F.S.; Investigation, A.U.R.K., M.R., M.W.M., M.M.F., H.H.S., M.J., M.F.S. and M.F.I.; Methodology, A.U.R.K., M.W.M., M.M.F., M.J. and M.F.I.; Project administration, M.R.; Resources, M.R., M.H. and H.H.S.; Software, M.R., M.H. and H.H.S.; Supervision, M.R.; Validation, A.U.R.K., M.W.M. and M.F.S.; Visualization, A.U.R.K., M.H., M.W.M. and M.M.F.; Writing—original draft, A.U.R.K.; Writing—review & editing, M.R., M.F.S. and M.F.I. All authors have read and agreed to the published version of the manuscript.

**Funding:** This work was supported by King Khalid University through a grant (KKU/RCAMS/22) under the Research Center for Advanced Materials Science (RCAMS) at King Khalid University, Saudi Arabia.

**Institutional Review Board Statement:** Not applicable.

**Informed Consent Statement:** Not applicable.

**Data Availability Statement:** The data related to this manuscript can be available on reasonable request from corresponding authors.

**Conflicts of Interest:** The authors have no conflict of interest regarding this manuscript.